\documentclass[10pt]{aastex}
\usepackage{emulateapj5}

\def\mj{M$_{\rm J}\ $}
\def\rj{R$_{\rm J}\ $}
\def\etal{{et~al.\,}}
\def\mo{M$_\odot$}
\def\ro{R$_\odot$}

\def\mp{M$_{\rm p}$}
\def\rp{R$_{\rm p}\,$}

\def\mstar{M$_{\ast}$}
\def\rstar{R$_{\ast}$}
\def\teff{T$_{\rm eff}\,$}
\def\teffs{T$_{\rm eff}$s$\,$}
\def\mic{$\mu$m$\,$}

\def\sles{\lower2pt\hbox{$\buildrel {\scriptstyle <}
   \over {\scriptstyle\sim}$}}
\def\sgreat{\lower2pt\hbox{$\buildrel {\scriptstyle >}
   \over {\scriptstyle\sim}$}}

\begin{document}

\slugcomment{\bf}
\slugcomment{Accepted to Ap.J.}

\title{A Theory for the Radius of the Transiting Giant Planet HD 209458b}

\author{Adam Burrows\altaffilmark{1}, D. Sudarsky\altaffilmark{1}
and W.B. Hubbard\altaffilmark{2}} 

\altaffiltext{1}{Department of Astronomy and Steward Observatory, 
                 The University of Arizona, Tucson, AZ \ 85721;
                 burrows@zenith.as.arizona.edu, sudarsky@as.arizona.edu}
\altaffiltext{2}{Department of Planetary Sciences, Lunar and Planetary Laboratory, 
                 The University of Arizona, Tucson, AZ \ 85721;
                 hubbard@lpl.arizona.edu}

\begin{abstract}

Using a full frequency-dependent atmosphere code that can incorporate
irradiation by a central primary star, we calculate self-consistent boundary conditions 
for the evolution of the radius of the transiting planet HD 209458b.
Using a well-tested extrasolar giant planet evolutionary code, we then
calculate the behavior of this planet's radius with age.  
The measured radius is in fact a transit radius that resides  
high in HD 209458b's inflated atmosphere: Using our derived atmospheric and interior structures, we find
that irradiation plus the proper interpretation of the transit radius
can yield a theoretical radius that is within the measured error bars.  
We conclude that if HD 209458b's true transit radius is
at the lower end of the measured range, an extra source of core heating power
is not necessary to explain the transit observations.

\end{abstract}


\section{Introduction}
\label{intro}

These past seven years have seen the number of known
extrasolar giant planets (EGPs) grow from 1 in 1995 (Mayor and Queloz 1995)
to more than 100 today.\footnote{see J. Schneider's Extrasolar Planet Encyclopaedia at http://www.obspm.fr/encycl/encycl.html
for an current tally, with ancillary stellar data.}  The wide variety in  
their projected masses (\mp sin(i)), orbital distances ($a$), and eccentricities continues to challenge
theories of EGP birth, evolution, and abundance. 
However, as of this writing, only two EGPs (HD 209458b and OGLE-TR-56b) are claimed  
to transit their primaries and of the two HD 209458b is by far the best studied. 
The second transiting EGP, OGLE-TR-56b, has only recently been suggested as such (Konacki \etal 2003)
and, at a distance of $\sim$1500 parsecs, even if the detection is verified, 
its light curve and radial-velocity measurements
are not yet competitive with those for HD 209458b.
Since HD 209458's stellar reflex motion has been accurately 
measured, and its transit light curve has been measured
using HST/STIS to $\sim$100-{\it micro}magnitude precision (Brown \etal 2001), it is an ideal 
testbed for the theory of irradiated EGPs, their evolution, structure, and atmospheres.
The transit of the F8V/G0V star HD 209458 lasts
$\sim$3 hours (out of a total period of 3.524738 days) and has an average photometric depth
of $\sim$1.6\% in the optical.  Its ingress and egress phases each last
$\sim$25 minutes (Henry \etal 2000; Charbonneau \etal 2000; 
Brown \etal 2001).  Using the complementary radial-velocity 
data (e.g., Henry \etal 2000), good estimates for the planet's mass,
orbital distance, and orbital inclination are $\sim$0.69 \mj, $\sim$0.045 AU, and $\sim$86$^{\circ}$, 
respectively.  Such proximity makes HD 209458b the quintessential ``roaster" 
(Sudarsky, Burrows, and Pinto 2000; Hubbard \etal 2001; Sudarsky, Burrows, and Hubeny 2002).
Of course, the depth of the transit light curve can be used to derive the planet's radius (\S\ref{data}).
It is the simultaneous availability of both a radius and a mass (together with
an age estimate for the system from stellar evolution theory and a luminosity estimate
for the star from its parallax) that makes this system 
especially useful to theorists.  Moreover, Charbonneau \etal (2002)
have recently binned their HST/STIS data to derive a wavelength dependence for HD 209458b's transit radius. 
In this way, they have inferred the presence of neutral sodium 
atoms (Seager and Sasselov 2000; Hubbard \etal 2001)
and, hence, have made the first measurement, however indirect, of the composition of the atmosphere of an extrasolar planet.

Importantly, the measured radius is a transit radius at a given
wavelength, or range of wavelengths.  It is not the canonical planetary radius at a
``1-bar" pressure level (Lindal et al.1981; Hubbard et al. 2001).  As such, the measured radius is the 
impact parameter of the transiting planet at which the optical depth to its primary's
light along a chord parallel to the star-planet line of centers is $\sim$1. This
is not the optical depth in the radial direction, nor is it associated with the radius at the 
radiative-convective boundary.   Hence, since the pressure level to which the transit beam is probing
near the planet's terminator is close to 1 {\it milli}bar (Fortney \etal 2003), 
there are many pressure scale heights
($\sim$10) between the measured transit radius and both the radiative-convective boundary 
($\ge$1000 bars) and the ``1-bar" radius.\footnote{If, as implied in Barman et al. (2002), the transit
radius is at pressures well below the 1 millibar level then the multiple-pressure-scale-height 
effect we identify here would be even larger.}
Furthermore, exterior to the radiative-convective boundary,
the entropy is an increasing function of radius.  One consequence 
of this fact is significant radial inflation
vis \`a vis a constant entropy atmosphere.  The upshot of both these effects 
is an increase of $\sim$0.1 \rj ($\sim$10\%) in the theoretical radius. 

Including the ``thickness of the atmosphere" could change recent interpretations of the 
radius of HD 209458b.  Guillot and Showman (2002), Baraffe et al. 
(2002,2003), and Bodenheimer et al. (2000,2003)
all have difficulty fitting HD 209458b's radius without an extra heat source.  Guillot and Showman
posit the dissipation at depth of mechanical energy generated by the stellar flux at altitude.
Bodenheimer et al. invoke tidal heating or the presence of an additional planetary 
companion in near resonance to create the ``Io" heating effect.  However, 
we find that self-consistently calculating irradiated
atmospheres and the irradiated planet's structural and thermal evolution and assuming that 
the lower end of the range of the measured transit radius obtains, we can fit
the observations without an extra source of heat.  Our calculations also self-consistently
and implicitly derive a Bond albedo, generally assumed by others as an input.   

In \S\ref{data}, we describe and discuss the measurements of 
HD 209458b's transit radius.  This is followed in \S\ref{previous} by
a brief review of previous theoretical calculations of the radii of irradiated EGPs.
In \S\ref{approach}, we summarize the computational procedures we employ 
in this paper and discuss the various strengths and weaknesses of our approach and in \S\ref{radius}
we present our theoretical results for the radius of HD 209458b, as well as its temporal evolution.   
This section contains the major conclusions of our study and includes in Fig. \ref{fig:2} what we
think is a favorable comparison between theory and measurement.  We wrap up in 
\S\ref{conclusion} with general remarks and caveats.

\section{The Measured Planetary Radius}
\label{data}

Photometric light curve measurements of the relative transit depth of the HD 209458 system from
both the ground (Henry \etal 2000; Charbonneau \etal 2000) and from space (HST/STIS: Brown \etal 2001)
have provided direct estimates of the ratio of the radii of the planet and star.  Mandel and Agol (2002)
conclude that this ratio can be obtained from the data to relatively high precision (\rp/R$_{\ast}$ = 0.1207$\pm 0.0003$).
Seager and Mall\'en-Ornelas (2003) suggest that future transit measurements alone, done with dense time sampling
and the best photometric precision attainable from the ground, can yield the individual radii themselves.
However, currently, estimates of the radius of the planet HD 209458b require estimates of the star's
radius and this requires both a fit to stellar evolution theory and a good parallax.\footnote{The Hipparcos distance
to the star HD 209458 is 47.3 parsecs, with an error of $\sim$5\% (Perryman 1997).}  
As a result, ambiguities in the stellar radius translate into uncertainties in the inferred
planetary radius.  Mazeh \etal (2000) conclude from $\log(g)$/\teff 
spectral-line fits and $M_{V}$/$(B-V)$ photometric fits that 
\mstar = $1.1\pm0.1$ \mo,  R$_{\ast}$ = $1.2\pm 0.1$ \ro, \teff$\sim$6000 K,
[Fe/H]$\sim$0.0, and $t = 5.5\pm 1.5$ Gyr.  From these data they derive a radius
for HD 209458b of 1.40$\pm0.17$ \rj, where the error bars are one-sigma and 
\rj = $7.149 \times 10^9$ cm.  Similarly, Cody and Sasselov (2002)
derive R$_{\ast}$ = $1.18$ \ro and \mstar = 1.06 \mo, with one-sigma error bars of $\sim$10\%. 
The resultant planetary radius is \rp = 1.42$^{+0.10}_{-0.13}$ \rj, where in this case the 
error bars in the planet radius do not include the errors in the stellar radius; Cody and Sasselov
assume in deriving this estimate that the stellar radius is fixed at 1.18 \ro. 
They also derive a best-fit age of $\sim$5.2 Gyrs ($\pm$10\%), but second the age range quoted in Mazeh \etal (2000).

Hence, to obtain a reliable stellar radius (given a parallax), one is obliged to derive the stellar metallicity,
age, mass, and helium fraction simultaneously.  This procedure perforce introduces ambiguities
into the estimate of the planetary radius.  In particular, if Cody and Sasselov were to use for the 
star's radius a value of 1.10 \ro (within their quoted errors), they 
would obtain a planetary radius of 1.32$^{+0.09}_{-0.11}$ \rj.
Hence, their one-sigma lower bound
would be 1.21 \rj.  At \rstar = 1.18 \ro, their one-sigma lower bound is 1.29 \rj.   
Similar arguments can be marshalled in the context of the Brown \etal (2001) and Mazeh \etal (2000)  
planet radius estimates.  As a result, it is not far-fetched to conclude that HD 209458b's transit radius
in the optical could be as small as $\sim$1.2 \rj.  While we are not completely convinced that this is the
correct value, it is nevertheless useful to explore the theoretical consequences of such a possibility.  
Given an age range for the HD 209458 system and a radius range for the planet, theories for the irradiated
planet's evolutionary trajectory in age-radius space can be compared with observations to derive
physical constraints on the nature of this irradiated planet and 
its atmosphere.

\section{Previous Theories for the Radii of Irradiated Giant Planets}
\label{previous}

The modern theory of the radii of irradiated EGPs begins with
the paper of Guillot \etal (1996).  This paper focussed on 51 Pegasi b, but 1) predicted that irradiation
due to proximity to a central primary would result in super-Jovian radii, 2) distinguished clearly
between the radii of hydrogen-rich and metal-rich giants (in the tradition of 
Zapolsky and Salpeter 1969), and 3) suggested that lower-mass
irradiated EGPs will be larger, all else being equal.  However, these 
calculations were performed using an irradiation-modified version of 
atmospheric boundary conditions (needed for the evolutionary calculations)
appropriate for isolated giants (Burrows et al. 1995; Burrows et al. 1997) and employed
an ad hoc Bond albedo.  The latter determines the absorption and heating efficiency of
stellar light.   A fully consistent spectrum and atmosphere calculation was not attempted.
Hence, while the results were qualitatively valid, the actual radius-age trajectories
obtained were ambiguous.  In addition, very large radii (1.4-1.8 \rj) seemed plausible
for older EGPs ($> 1$ Gyr).  

In the first year of the HD 209458b campaign, Burrows \etal (2000) employed
similar boundary algorithms and Bond albedo ansatze to fit HD 209458b's measured transit radius.
They concluded that the planet could not be as large as measured unless it had
migrated in early in its life before it had had time to cool and shrink appreciably.  They
showed that once an EGP achieves mere Jovian proportions it cannot be inflated
enough to conform to the new transit radius measurements.  While we concur with this
general conclusion, we disagree with the magnitude of the discrepancy that late migration
would have created.  If, as we suggest in \S\ref{data}, HD 209458b's transit radius can 
be as small as $\sim$1.2 \rj, we now find that the discrepancy in the radius would be $\sim$0.1 \rj,
not $\sim$0.3 \rj, as implied in Burrows \etal (2000).  This is still significant, but less so.
This modified conclusion is a consequence of our better and more consistent boundary
conditions and the use of a state-of-the-art, frequency-dependent stellar atmosphere code that can
accurately handle insolation (\S\ref{approach},\S\ref{radius}).

Recently, Guillot and Showman (2002), using a modified version of the Guillot \etal (1996)
boundary conditions, have questioned the larger radii derived in Guillot \etal (1996) and 
Burrows \etal (2000).  They conclude that HD 209458b's radius cannot be explained without an extra
heat source and they speculate that this might be due to the degradation at depth of gravity waves generated
in the upper atmosphere by a fraction of the stellar flux.  However, in order to obtain boundary
conditions for their new irradiated-planet evolutionary calculations (their ``cold" case, 
without an extra heat source), they modify the atmospheric temperature/pressure
profiles of isolated models (Burrows \etal 1997) by shifting the temperature at 3 bars by a  
somewhat arbitrary 1000 Kelvin.  They suggest that this procedure mimics the effect
of the thick radiative zones of close-in EGPs, through which stellar light can
not penetrate to heat the convective interior.  While no attempt is made to construct self-consistent
atmospheres that incorporate actual stellar and planetary spectra, this procedure does qualitatively capture
the expected differences in T/P profiles between distant Jovian planets and nearby roasters at the same
gravity and interior entropy.  The result is an HD 209458b model with a radius no greater than $\sim$1.1 \rj.  
Guillot and Showman (2002) have a discussion of the possible differences 
between day-side and night-side cooling (see \S\ref{approach}), but do not provide 
definitive quantitative guidance concerning this important, and still open, issue.  

The results of Bodenheimer, Lin, and Mardling (2001) and Bodenheimer, Laughlin, and Lin (2003),
who assume a Bond albedo and that the atmospheric temperature at $\tau_{Rosseland} = 2/3$ is equal to \teff,
are quantitatively similar to those of Guillot and Showman (2002) and they explore the possible effects 
of heating by tidal circularization and forcing by a second planet or of the explicit 
presence or absence of a heavy-element core.  However, recently,
in parallel with the present work, Baraffe \etal (2002,2003) have begun to incorporate more realistic atmospheric
boundary conditions into HD 209458b studies using a frequency-dependent atmosphere code. 
Nevertheless, they too obtain smaller radii than found in  
the Guillot  \etal (1996) and Burrows \etal (2000) studies and, as a consequence, also evoke an another
source of heat to explain the measured transit radius.    

It is the thesis of this paper that a combination of the proper interpretation of the 
transit radius (\S\ref{intro}) and a lower measured value of that radius (still within the 
error bars, \S\ref{data}) can be shown to be in accord with consistent and 
frequency-dependent atmosphere and evolutionary calculations, without an additional source of heat.  
These calculations do not use an ad hoc Bond albedo and do use a realistic G0 V irradiation spectrum. 
However, the completeness of our theory as an explanation for 
HD 209458b's transit radius hinges upon the assumption that the true radius
resides at the lower end of its measured range.  We now turn to a description of our computational
procedures.

\section{Computational Methods and Assumptions}
\label{approach}

At the low effective temperatures achieved by brown dwarfs and EGPs, 
boundary conditions for evolutionary calculations must 
incorporate realistic atmospheres (Burrows \etal 1997; Allard \etal 1997).
The traditional method of setting the effective temperature equal to the temperature
at a Rosseland mean optical depth of $2/3$ does not really provide the T/P
profile in the atmosphere and can lead to errors in \teff of hundreds of Kelvin. 
Hence, for isolated objects with cold molecular atmospheres, we calculate 
a grid of detailed T/P profiles and spectra at various \teffs and gravities ($g$).  These atmospheres penetrate 
deeply into the convective core.  In this way, we determine the relationship between
the core entropy ($S$), \teff, and $g$ (Hubbard 1977).  Since the core contains the mass and the
heat, while the thin atmosphere is the valve that regulates radiative losses,
interpolating in this pre-calculated $S$-\teff{-$g$} grid at each timestep in an
evolutionary calculation assures accuracy and self-consistency.     

However, when an EGP is being irradiated by a primary star, the above procedure must
be modified to include the outer stellar flux in the spectrum/atmosphere calculation that yields
the corresponding $S$-\teff{-$g$} relationship.  This must be done for a given external stellar
flux and spectrum, which in turn depends upon the stellar luminosity spectrum and the orbital distance of the 
EGP.   Therefore, we use the Discontinuous Finite Element (DFE) variant of the spectral code
TLUSTY (Hubeny 1988; Hubeny and Lanz 1995) that we have developed
for EGP and brown dwarf atmospheres (Sudarsky, Burrows, and Hubeny 2002) 
to calculate a new $S$-\teff{-$g$} grid under the irradiation
regime of HD 209458b.  This grid is tailor-made for the luminosity and spectrum of HD 209458 at a distance
of 0.045 AU.  For the stellar spectrum, we employ a theoretical G0 V spectrum of Kurucz (1994)
and we assume that the orbital distance, stellar luminosity, and stellar 
spectrum are all constant during the EGP's evolution.  (Hence, we ignore the ``faint young sun"
problem.)

The procedure is straightforward.  For a given inner flux boundary condition (represented by \teff,
where $\sigma {\rm T}^4_{\rm eff}$ is the interior core flux of the EGP) and a given gravity $g$, we calculate
the atmospheric T/P profile, including the irradiation.  Using mixing-length theory, the atmosphere is continued 
deep into the convective region.  The radiative-convective boundary may be at pressures 
of $\sim$10 bars (for younger ages) to $\sim$4000 bars (for older ages).
The entropy of the core ($S$) is now known for a given interior heat flux (parametrized by \teff) and gravity ($g$).  
After the table in $S$, \teff(internal flux), and $g$ is generated, it is inverted to obtain the more
useful relationship \teff($S$,$g$) for the interior flux.  This is the function used to advance the evolutionary
calculations for HD 209458b, using a state-of-the-art equation of state (Saumon, Chabrier, and Van Horn 1995)
and evolutionary code (Burrows \etal 1997).  Note that since an EGP core 
is convective the radius of the radiative-convective boundary
for a given mass EGP (e.g., 0.69 \mj) is a function of $S$ and $g$ alone.  Implicit in this procedure is an albedo, which therefore
does not have to be assumed, as well as an emergent planetary spectrum and a T/P profile to pressures
below a microbar. Since we calculate the atmospheric profile, we can automatically include the 
multiple-scale-height effect described in \S\ref{intro} in our calculation of the transit radius.  

We have explored the effects on the resultant EGP transit radius of 
variations in the helium fraction (Y$_{He}$), of the presence or absence
of clouds in the upper atmosphere (Sudarsky, Burrows, and Hubeny 2002), 
of the possible presence of a rocky core, and of changes in the opacities in the 
deep atmosphere (e.g., due to the presence or absence of TiO/VO).  
Note with the latter that our motivation for varying the opacities  
in this manner is not to suggest that TiO or VO may not be present (we certainly think they are),
but to explore thereby the dependence on the radius evolution of large changes
in the opacities at depth.  Increasing Y$_{He}$ from 0.25 or 0.28 to 0.30
can mimic the effect of a metal-rich ``rock" core or envelope and slightly 
shrink the object (Zapolsky and Salpeter 1969; Guillot and Showman 2002).  For instance, 
we have calculated that for an EGP with a mass of 0.69 \mj replacing a one-Earth-mass heavy-element
core by one containing 11 Earth masses can shrink the planet's radius by 3-4\%.
This amounts to approximately -350 kilometers per Earth mass. 
An increase in the core mass by 10 Earth masses is comparable to the effect of increasing Y$_{He}$ 
by $\sim$0.02 units.  The models plotted in Figs. \ref{fig:1} and \ref{fig:2} do
not incorporate a rocky core, though some of the models assume a high Y$_{He}$ to mimic
its presence.   However, there are two major
issues that in the final analysis will need to be addressed in detail before a 
definitive answer is derived.  For the time being, they must
be finessed and have been finessed in all previous theoretical studies.  The first is the fact that
the planet is spherical, while the insolating stellar flux is a bundle of parallel beams.  This results
in a flux whose angle of incidence is a function of latitude.  At the sub-stellar point the 
flux is a maximum, while it decreases as the terminator is approached.   
We have introduced the flux parameter $f$ which accounts in approximate fashion for the 
variation in incident flux with latitude when using a planar atmosphere code such as we employ. 
Hence, a value of $f = 1/2$ is a reasonable average and is our baseline value, but we have explored the 
consequences of the bounding values $f =1.0$ (appropriate for the sub-stellar point) and $f= 1/4$.  

The second major issue is the day-night
cooling difference.  The gravity and interior entropy are the same for the day and the
night sides of HD 209458b.  For a sychronously rotating planet such as HD 209458b,
the higher core entropies needed to explain a large measured radius imply higher internal fluxes
on a night side if the day and the night atmospheres are not coupled (Guillot and Showman 2002).  As described in
Burrows \etal (2000), the day side core flux is quite suppressed by the flattening of
the temperature gradient and the thickening of the radiative zone due to irradiation.
However, Showman and Guillot (2002), Menou \etal (2002), and Cho \etal (2003) have recently demonstrated
that strong atmospheric circulation currents that advect heat from the day to the night
side at a wide range of pressure levels are expected in HD 209458b.  Showman and Guillot (2002)
estimate that below pressures of $\sim$1 bar the night-side cooling of the air
can be quicker than the time it takes the winds to traverse the night side, but that at higher pressures
the cooling timescale is far longer.  Importantly, the radiative-convective boundary
in HD 209458b is deep in the planet, at pressures above 1000 bars.  We take this to mean that
due to the coupling of the day and the night sides via strong winds at depth, the temperature-pressure
profiles at the convective boundary on both sides are similar (A. Showman, private communication).  
This implies that HD 209458b's core cooling rate is roughly the same in both hemispheres.  
This is similar to the case of Jupiter,
where the interior flux is latitude- and longitude-independent, despite solar irradiation.  
Clearly, a full three-dimensional
radiation/hydrodynamic study or Global-Climate-Model (GCM) is necessary to resolve this 
thorny issue definitively, but in the interim we assume in this 
paper that the interiors on the day and night sides cool
at the same rate.   

In calculating our T/P profiles, we assume a constant $g$.  Since $g$ in fact decreases slightly
from the bottom to the top of HD 209458b's extended atmosphere, the thicknesses of our calculated
atmospheres are slightly underestimated.  Furthermore, winds carrying gas from the day side to the 
night side (and vice versa) should experience temperature changes at altitude that might alter the composition
and cloud profiles.  Since a transit observation probes the day-night terminator, such variations
could affect the character of the transit itself.  However, for this study, we ignore this potential
complication. We use the opacity library described in Sudarsky, Burrows, and Hubeny (2002)
and 5000 frequency points from 0.4 to 300 \mic in calculating the T/P profiles.

\section{Results: R$_p$ versus Age}
\label{radius}

\begin{figure}
\epsscale{1.00}
\vspace*{-0.7in}
\plotone{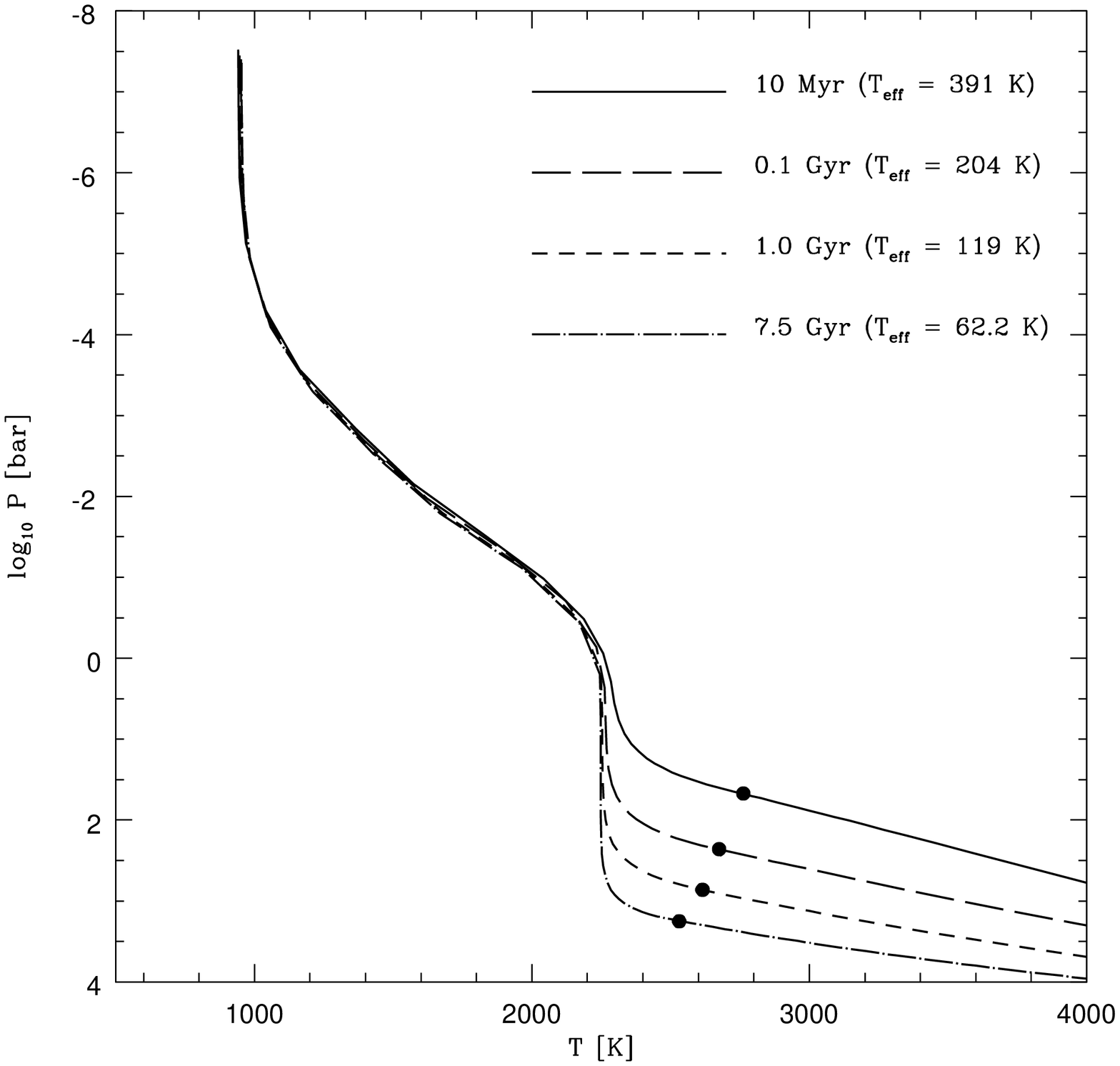}
\vspace*{-0.2in}
\caption{
Temperature (in Kelvin) versus pressure (in bars) profiles along the evolutionary
track of the model depicted in solid blue on Fig. \ref{fig:2}.  The parameters
for that curve are $f =0.5$ and Y$_{He}$ = 0.30 and the TiO/VO opacity at depth has been dropped.
The large dots indicate the positions of the radiative-convective boundary.  The solid curve
is at an age of 10 Myr and has a \teff (inner boundary flux) of 391 K, the long dashed curve
is at 0.1 Gyr and has a \teff of 204 K, the short dashed curve is at 1.0 Gyr and has
a \teff of 119 K, and the dot-dashed curve is at 7.5 Gyr and has a \teff of 62.2 K.
See Fig. \ref{fig:2} and the text for details.
\label{fig:1}}
\end{figure}

\begin{figure}
\epsscale{1.00}
\vspace*{-0.7in}
\plotone{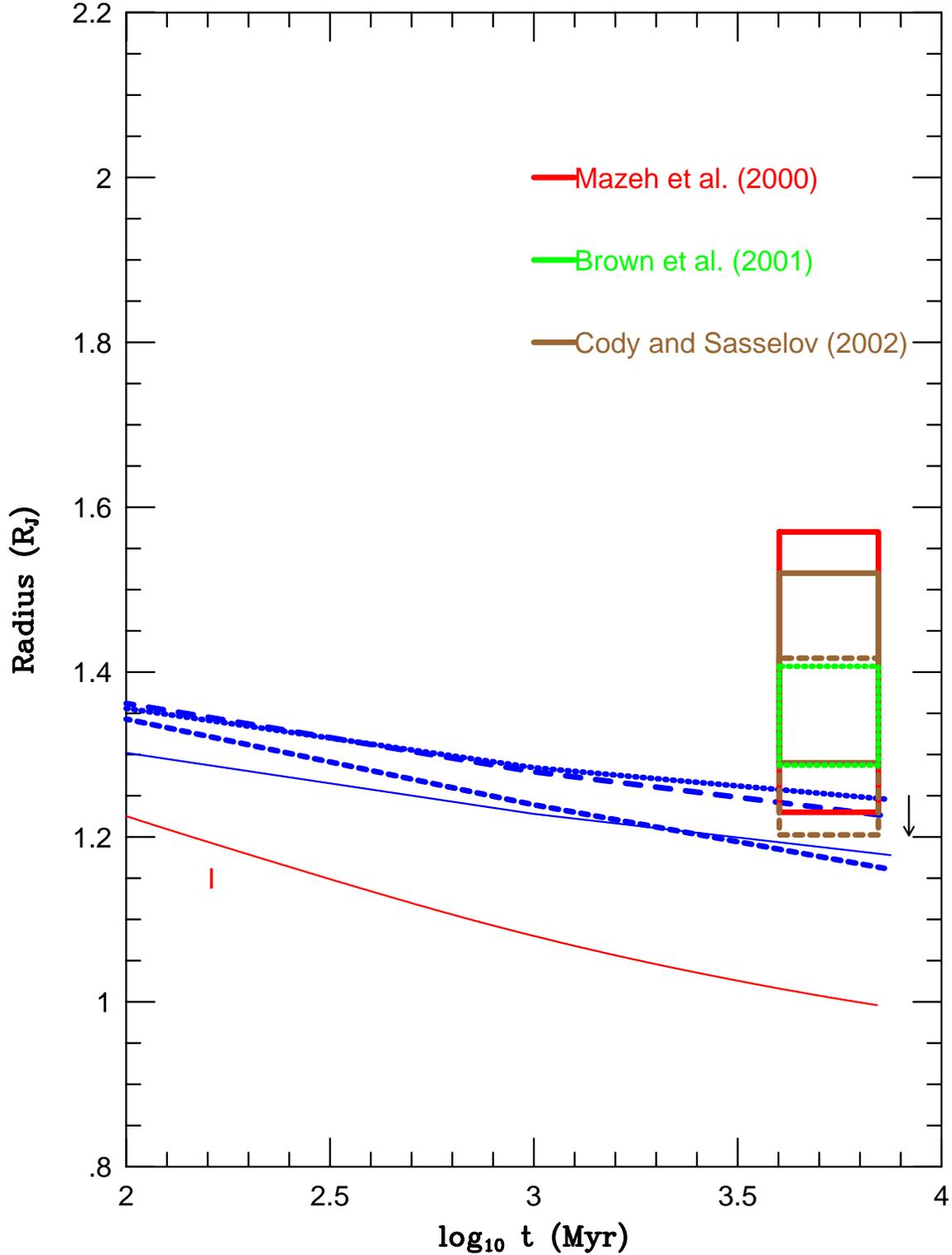}
\vspace*{-0.8in}
\caption{
Theoretical evolutionary trajectories (blue) of the radii of
HD 209458b (in \rj; $7.149\times 10^4$ km) with age (in Myrs).
Model {\bf I} (solid) is for a 0.69 \mj object in isolation (Burrows \etal 1997).  The $\pm 1-\sigma$ error
boxes for the inferred transit radii from Mazeh \etal (2000), Brown \etal (2001), and Cody and Sasselov (2002)
are shown in red, green, and gold, respectively.  The solid gold box assumes a stellar radius of 1.18 \ro,
while the dashed gold box assumes a stellar radius of 1.10 \ro. The solid blue curve is for $f =0.5$, Y$_{He}$ = 0.30,
does not include TiO/VO opacity at depth, and does
not have clouds (see Fig. \ref{fig:1}).  The short dashed blue curve is similar, but includes
both TiO/VO opacity and a forsterite cloud at altitude (Fortney \etal 2003).
The dotted blue curve is for $f = 1.0$, but otherwise has
the same parameters as the solid blue curve, and the long-dashed blue
curve has no TiO or VO at depth, no cloud, and a helium fraction
of 0.25.  The short arrow to the right of the error boxes depicts the magnitude
of the radius decrease for each 10 Earth-mass increase in the mass of
a possible rocky core, all else being equal.
See text for details.
\label{fig:2}}
\end{figure}

Figure \ref{fig:1} portrays temperature-pressure profiles at various ages along the solid model
trajectory plotted in Fig. \ref{fig:2} with $f =0.5$, Y$_{He}$ = 0.30, and without TiO/VO opacities.
We plot the T/P profiles for this model only as one example among many. 
The positions of the radiative-convective boundaries are indicated with large dots and 
each profile is for a specific time (and \teff) during the evolution.  
The pronounced inflection in the curves is predominantly
a result of the near balance at some depth between countervailing incident and internal fluxes,
but changes in the opacity profiles also play a role.  Nevertheless, such inflections are
generic features of consistent calculations of irradiated atmospheres and affect the mapping
between \teff, $S$, and $g$ used in evolutionary calculations.  As is indicated on Fig. \ref{fig:1},
the \teff for this model at 7.5 Gyr is near $\sim$62 K, which implies a $\sim$65\% lower internal flux than obtained
for the isolated ({\bf I}) model.  For the isolated model at 7.5 Gyr, \teff is $\sim$82 K.
The large contrast between these low \teffs and the high atmospheric temperatures (1000-2000 K)
seen in Fig. \ref{fig:1} for an irradiated object may seem 
counterintuitive, until one realizes that \teff is the interior ``flux temperature" from
which the core luminosity is derived.  Evolution over time and the stanching of heat
loss by the hot atmospheric thermal blanket result in flux temperatures for irradiated
objects that are smaller than, but of the same order as, those for isolated or orbitally 
distant EGPs (Burrows et al. 2000).

Figure \ref{fig:2} compares the transit radius data summarized
in \S\ref{data} with various representative radius-age trajectories calculated in this paper.
The theoretical ages start at 10$^8$ years and the radii are in Jovian units (1 \rj $\equiv$ 7.149$\times 10^4$ km).
As described in \S\ref{intro}, the radius at the 1.0 millibar level is taken to be the
relevant radius (Fortney \etal 2003).  In this way, the thickness of the atmosphere, which thereby contributes 
0.08 \rj to 0.13 \rj to the total radius, is incorporated into the theoretical numbers.
The 1-$\sigma$ radius ranges inferred by Mazeh \etal (2000, red), Brown \etal (2001, green), and Cody and Sasselov (2002, gold)
are plotted; for all observational estimates the age is assumed to be 
bracketed by 4 and 7 Gyr.  For the Cody and Sasselov (2002) data, two boxes, one assuming a stellar radius
of 1.18 \ro (solid gold) and the other assuming a stellar radius of 1.1 \ro (dashed gold), are provided. 
The red curve adjacent to an {\bf I} is the corresponding trajectory for an isolated EGP with 
a helium fraction of 0.28.  As one can clearly see, the difference between the {\bf I} trajectory and the irradiated
trajectories is approximately $\sim$0.2 \rj.

The theoretical irradiated trajectories are all shown in Fig. \ref{fig:2} 
in blue.  The solid curve is for $f =0.5$, Y$_{He}$ = 0.30,
does not include TiO and VO at depth, and does 
not have clouds (see Fig. \ref{fig:1}).  The short dashed curve is similar, but includes
both TiO/VO at depth and a forsterite cloud at altitude 
(Fortney \etal 2003).  The dotted curve is for $f = 1.0$, but otherwise has 
the same parameters as the solid curve, and the long-dashed 
curve has no TiO or VO at depth, no cloud, and a helium fraction
of 0.25.   The presence or absence of TiO/VO is meant to gauge the effect of significant changes in
opacity at higher pressures and temperatures.  We see that the effect of 
altering this opacity, while discernable, is not large.  
This representative model set illustrates and quantifies several clear systematic trends.  The first
is that the radius is a decreasing function of increasing Y$_{He}$ (or metallicity).  A larger Y$_{He}$ might
mimic the effect of a rocky core (Guillot and Showman 2002).  The second is that
a larger $f$ yields a larger radius.   Intense irradiation inflates the atmosphere and at a given epoch results in  
interiors with slightly higher entropies.  The third is that the extra opacity effect of a cloud seems to decrease
the radius.  However, this last trend is actually a bit misleading, since the presence of a cloud also puts the transit
radius at lower pressure levels, thereby increasing the measured radius.  The two effects roughly cancel.  
For none of the models shown is there an extra source of heating power.  The finite thickness of the atmosphere 
and the low pressure level of the actual transit radius account for $\sim$0.1 \rj of our theoretical radii and 
are the major reasons we differ from the theories of Guillot and Showman (2002), Bodenheimer, Lin, and Mardling (2001),
and Bodenheimer, Laughlin, and Lin (2003). Baraffe et al. (2003) do account for the finite thickness of
the atmosphere, but their estimated thickness is about
half the thickness that we obtain here.  Moreover,
details of the atmospheric boundary condition govern
the core entropy, and, hence, the overall size of the planet.
This dependence can also lead to dispersion in theoretical
estimates of the planet's size.  

The general proximity of the sheaf of 
models on Fig. \ref{fig:2} to the 1-$\sigma$ lower bounds of the 
measured radius is what motivates us to suggest that an additional heat source may not be necessary to explain
the HD 209458b radius measurements.  However, the effective Y$_{He}$ or $f$ and the role of day-night
thermal coupling by winds are not yet sufficiently well constrained to pinpoint the theoretical
radius of the 0.69-\mj EGP, HD 209459b, to better than $\sim$0.05-0.1 \rj.  Note that we obtain a theoretical upper
bound to the transit radius of HD 209458b at 4.0 Gyr of $\sim$1.3 \rj 
if Y$_{He}$ = 0.25, $f$=1.0, there is no rocky core,
and the TiO/VO opacity is suppressed.  Radii greater than this for 
HD 209458b will be difficult to explain without significant
alterations in the physics or chemistry and/or without an extra source of internal power.

\section{Conclusions}
\label{conclusion}

We have calculated consistent T/P profiles for various models of an irradiated HD 209458b.
These profiles penetrate deeply into its inner convective zone and have been used to derive
realistic boundary conditions for evolutionary models of its transit radius.  The latter
has been pegged to the $\sim$1-millibar pressure level inferred in the work of Fortney \etal (2003) and
Hubbard \etal (2001) and accounts for $\sim$10\% of the total radius.  We find that irradiation 
and the proper interpretation of the measured radius as the transit radius alone can explain 
HD 209458b's observed radius, without invoking an additional source of heating, if its true 
transit radius is at the lower end of the measured range.   

Major caveats and uncertainties remain and include the actual effect of heat transport by winds from the day to the
night sides, the possible presence of a rocky core, the actual helium fraction, and 
the true effective insolation parameter $f$. To resolve most of these issues will require
multi-dimensional modeling with realistic radiative/convective transport.  
If the actual helium fraction is greater than $\sim$0.28, then
a rocky core much larger than $\sim$15 Earth masses may be difficult to accommodate.  If better measurements
of the transit radius indeed yield large radii for HD 209458b at or above $\sim$1.40 \rj, then our interpretation
is not tenable.  However, if radii closer to the current lower error bars are proven to obtain,
we would conclude that the theory of the structure of irradiated EGPs is in hand
and well poised to interpret the new transiting EGPs anticipated to be discovered in the next few years.

Curiously, the preliminary radius for OGLE-TR-56b obtained by Konacki \etal 2003) is
only $1.3\pm$0.15 \rj, despite its extreme proximity ($\sim$0.023 AU) to its central sun-like star.
Having a mass ($\sim$0.9 \mj) similar to that of HD209458b ($\sim$0.69 \mj), one might have expected
the radius of HD209459b (at an orbital ditance of $\sim$0.045 AU) to have been 
smaller, not larger, than that of OGLE-TR-56b.  This could suggest that our theory
for irradiated EGPs is basically correct, and that additional heat sources are not
the norm, but that the transit radius of HD209458b is anomalously large, perhaps because
of the special circumstance of a companion (as suggested by Bodenheimer et al.).
With a collection of transits, supplemented with good radial-velocity data, and 
a family of radius-mass pairs for different irradiation regimes, we will be able
to check this hypothesis, as well as our general theory for irradiated planet transit radii.   
Such a family will be a major boon to the emerging study of extrasolar giant planets.

\acknowledgments

We would like to thank Ivan Hubeny, Dimitar Sasselov, Adam Showman, Jonathan 
Lunine, Drew Milsom, Jonathan Fortney, Curtis Cooper,
and Christopher Sharp for insightful discussions 
during the course of this work. This work was supported 
in part by NASA grants NAG5-10760 and NAG5-10629. 

{}

\clearpage

\end{document}